# Quantum Technology master's: A shortcut to the quantum industry?

*Simon Goorney ([1,2,*]), Borja Muñoz ([1,2,&]), Jacob Sherson ([1,2,**])*

*[1]: European Quantum Readiness Centre*

*[2]: Department of Management, School of Business and Social Science, Aarhus University*

*Corresponding authors*

*simon.goorney@mgmt.au.dk

** sherson@mgmt.au.dk

## ORCID

*Simon Goorney:* https://orcid.org/0000-0002-8635-7523

*Jacob Sherson:* https://orcid.org/0000-0001-6048-587X

## Twitter

*Simon Goorney* @simongoorney

*Jacob Sherson:* @jacobsherson

## Abstract

In this article, we investigate a growing trend in the worldwide Quantum Technology (QT) education landscape, that of the development of master's programs, intended to provide graduates with the knowledge and skills to take a job in the quantum industry, while serving a much shorter timeline than a doctoral degree. Through a global survey, we identified 86 master's programs, with substantial growth since 2021. Over time master's have become increasingly interdisciplinary, organised by multiple faculties or through joint degree programs, and offer more hands-on experiences such as internships in companies. Information from program organisers and websites suggests that the intended career destinations of their graduates are a diverse range of industries, and therefore master's programs may be a boon to the industrialisation of quantum technologies. Finally, we identify a range of national efforts to grow the quantum workforce of many countries, "quantum program enhancements", which augment the content of existing study programs with quantum content. This may further contribute to the growth and viability of master's programs as a route to the quantum industry.

# 1.Introduction

The second quantum revolution has opened a substantial array of opportunities and challenges that come with the materialization of new technologies based on the laws of Quantum physics. The so-called “quantum industry”, defined as “companies that use QIST [Quantum Information Science and Technology] in their business/products or provide technologies that enable such business/products”[1], is growing and rapidly specialising. As a result, the need to supply the demand for workers is becoming more evident[2], and understanding the quantum educational landscape can provide valuable insights into the progress in this direction worldwide. In this respect, several recent studies have addressed the question of what qualifications, training and skills are needed to enter quantum industry.[3,4]

A concrete example of the different competencies that could be used as a reference in the development of educational programs is the European Competence Framework for Quantum Technologies[3]. Existing efforts to standardise the future quantum workforce are already being implemented and have indicated, for example, levels of proficiency based on expected knowledge and skills[3] aligned with industry needs. Following this, several studies have also presented concrete guidelines that could be followed for creating educational programs regarding quantum, including bachelor's and master's programs.[5,6,7,10]

## 1.1 The needs of the Quantum Industry

Exactly what are, and will be, the needs of the industry are important questions because they should inform educational practices. The educational requirements of different types of position in the quantum industry was first investigated qualitatively in 2020 by Hughes et al[1]. when they surveyed 57 companies on the important knowledge and skills for key roles in their organisations. The distribution was rather striking in the number of jobs, even rather technical ones, for which companies did not require a PhD graduate to fulfil, instead quoting that a master's graduate could fill the position. For example, with respect to the position "Cryogenics Engineer", roughly 60% of the surveyed companies stated a master's graduate could fill this role. The authors note that the most highly specialised roles required a higher fraction of PhD graduates. However, as the quantum industry is continually diversifying into new markets and application domains such as healthcare, transportation, finance, and many others, we would expect the proportion of application-level positions to increase, and thus the number of roles accessible to master graduates could increase. For this reason, master's programs could be a key route into the quantum industry, offering "specific quantum expertise"[2] without the significant timeline associated with PhD programs.

Fox et al[4] discussed that despite the prevalence of physics departments in preparing students for the quantum industry, engineers are needed for the realisation of quantum products. In terms of the skills needed to enter the quantum industry, it is important to note that "classical skills" in physics and engineering are highly valued by the quantum industry, particularly in the hardware development sector. The authors suggest that, as of 2020, quantum-relevant knowledge was missing from engineering programs, and that perhaps master's degrees could provide this for engineering graduates with a non-quantum background[3]. Aiello et al[8] and Franziska et al[4] reached the same conclusion, arguing that Quantum Information Science and Engineering (QISE) ought to be included in all STEM major degrees in order to promote widespread quantum literacy.

Although quantum mechanics within undergraduate programs has traditionally been associated with physics departments, authors such as Asfaw et al.[5] stress the importance of expanding to other fields, including but not limited to applied mathematics, chemistry, computer science, electrical engineering, materials engineering, and molecular engineering, therefore aligning with its cross-disciplinary and demands of the quantum workforce. In their investigation in 2021, Aiello et al[8] identified that among faculty teaching a sample of quantum masters, Physics departments were predominant (41%) followed by 35% in multiple departments, and only 18% hosted by Engineering departments.

Following the analysis of course catalogues from 305 public institutions in the United States by Cervantes et al.[9], it was shown that although a fairly large number of QIS courses were offered by the departments of Electrical and Computer Engineering (ECE) and Computer Science (CS), fields heavily impacted by quantum computing, Physics departments had the highest number of QIS courses compared to other departments. The authors suggest It would be beneficial to recognize the interdisciplinary nature of QIS to enhance the quality of education in this field, implementing cross-departmental collaboration for the progression of QIS education.

Several authors[1,2,8] have noted that hands-on experience is of crucial importance in offering the “specific quantum expertise”[2]. desired by industry. Investigating this, Aiello et al[8]. addressed a sample of master’s Quantum Science & Engineering programs to examine if and how they include hands-on experiences. They analysed 14 masters programs for the inclusion of lab experience or physical demonstrations, internships with industrial partners, or projects and theses related to research. Of these, only 5 offered the opportunity for students to spend time working with an industrial partner, which may be far from ideal given that such opportunities could be beneficial in granting access to the job market for graduates. The relatively small sampling, and the pace at which quantum technologies are evolving, calls for a more in-depth look into what experiences are offered by quantum master’s programs, as we address in this article.

Equally important, the student perspective regarding QT and quantum careers has been recently analysed in undergraduate STEM disciplines. A recent analysis revealed that most of the students knew “nothing” about quantum careers (55%), and as a result of the focus groups conducted, student’s perceived that quantum is only for “geniuses” or those with an advanced education, such as obtaining a PhD[11]. In this paper, we examine if that is indeed the case, or whether master’s programs could provide an alternative route to the quantum industry.

## 1.2 Workforce development in National Quantum Strategies

Many nations of the world have therefore addressed the importance of master's programs by emphasising them in their National Quantum Strategies, including Europe[12],[13], United States[14], Israel[15], South Korea[16], Australia[17], France[18], Switzerland[19], South Africa[20], Canada[21], Ireland[22], Slovakia[23], Denmark[24], Netherlands[25], Thailand[26], Germany[27]. In particular, South Africa stresses the importance of sharing curricula in "disciplines ranging from physics, to engineering, and computer science". In addition, specialised programmes at Masters level can be introduced to "accelerate the research training and address the needs of industry"[20]. Another example is Switzerland, which highlights the need for interdisciplinary education as a "collaborative educational project including federal institutes, universities and if possible the UASs [universities of applied sciences]"[19]. Among its key actions, Ireland stresses the need to "share resources and expertise, e.g., through consortia for master's and doctoral training, and increase the scale of quantum-related education as part of the learning offerings at higher-education institutions (HEIs)"[22].

In Europe, attention has gone toward offering the quantum expertise to graduates of non-quantum master's programs, through experimentation with an "Open Master" program as part of the European QTEdu community[28],[29],[30]. There are other efforts worldwide which represent a significant upscaling in quantum education at the national level, an emerging trend which we will examine further in the discussion section.

In this article, we investigate the landscape of master's programs in Quantum Technology. How have the number of master's programs worldwide evolved over time, and do they reflect the increasing diversification of industries in which quantum skills are applied? And is studying for a Quantum Master a viable route to an industry career, as opposed to the more traditional option of doctoral studies?

# 2. Methodology

## 2.1 Criteria for a Quantum Technology master's program

We have recognised a Quantum Technology Master as a one or two-year program awarding a master's degree and specialising in at least one of quantum computing, quantum sensing, quantum communication, quantum simulation, or engineering devices based on these technologies. Of all quantum technology master's, a further distinction has been established between **primary master's** and **secondary master's**. **Primary master's** are those in which the entire program is structured around at least one of quantum computing, quantum sensing, quantum communication, quantum simulation, or quantum engineering, and this is reflected in the degree title. **Secondary master's** are those which are quantum technology specialisation tracks or concentrations of non-specialist master's degrees such as Physics.

These definitions exclude master´s degrees specialising in, for example, nuclear engineering, nanotechnology or photonics. Whilst these are indeed technologies which make use of quantum mechanical properties, they are not commonly considered "quantum technologies" as the term is currently used to describe technologies arising in the "second quantum revolution"[31]. We also have excluded masters longer than 2 years of duration, for example, integrated bachelor’s and master's degrees, of which several exist in the UK[32,33].

## 2.2 Searching for and identifying programs

The process of identifying these master’s programs has involved the use of existing, non-exhaustive lists available in some sources[28,34] updated by searching the internet by the country for master’s programs. To ensure the validity of the programs, information was verified with the program coordinators or representatives of each master's program through email. Therefore, we have gathered information regarding the country, city, name of the program, link to their website, the first entry of students to the program, faculty or faculties responsible for the program, whether they offer internships and the expected career outcomes of the students.

## 2.3 Investigating hands-on experience: Internships offered in master programs

In extending and updating the work of Aiello et al[8] investigating hands-on experience offered in master’s programs, we studied 51 primary master’s to determine the type of internship they offer. To do so we gathered 51 internship offers as described by program websites and/or communication with faculty representatives, and divided them into three categories: "Internship university" pertains to university courses that involve projects, case studies, or internships offered by academic research groups or laboratories affiliated within the university offering the master program. “Internship industry” refers to internships carried out in industrial settings including projects which involve collaboration with companies (on-site or online). Finally, there is a category for those internships which are flexibly offered either in the university or industrial settings, providing the possibility of choosing towards industrial or academic specialisation.

## 2.4 Faculties organising the programs

A further important feature to understand in the growth of the landscape of quantum technology education is its increasing interdisciplinarity. To provide a clear understanding of the faculties running the master's programs, we have clustered them into four main categories: Physics, Engineering, Computer Science and Joint faculties. If a master program involves the participation of multiple faculties, we have categorised it as a Joint faculty. On the other hand, if the program is more closely aligned with a specific faculty, we have categorised it as such.

## 2.5 Assigning graduate career outcomes

To obtain the information regarding career outcomes, we used two distinct data sources in order to provide as rich and detailed a description as possible. The first of these **(D1)** was a survey sent to program coordinators in which they selected a list of potential industries/sectors in which students could find a job after their studies. The list was generated by adapting standardised industry data used by one of the largest job boards online[35], and used by the EU's Quantum Readiness Centre[36] to provide comparison between job offers and career outcomes in future work.

The second data source **(D2)** were the program websites themselves, whereby text referring to projected career outcomes for graduates was extracted. This text was transformed into txt format, resulting in 46 documents belonging to different universities. We used Voyant tools[37] to upload the 46 documents which contained a total of 721 unique words (each term is listed once despite having multiple occurrences) and 2583 occurrences. To refine our dataset, we defined a customised list of stopwords[38] that appeared frequently in our corpus but did not hold significance in relation to career destinations. This involved, but was not limited to, removing words like, prepositions, articles, adverbs as well as verbs denoting actions such as "research", "programming", or "teaching", as they refer to intended competencies or learning outcomes rather than career outcomes described in terms of industries.

Hence**,** after removing the stopwords (a total of 631 unique stopwords and 2322 occurrences) we identified 90 words and studied in detail each individual occurrence of these words in context (a total of 241 occurrences) to confirm that they were indeed referring to industries. To avoid overclassification within a single document, if a word was already assigned to a career destination (e.g: if the word "Phd" was categorised as "Academic research"), we avoided assigning a similar word in the same document to the same category (e.g: the word "Academia" as "Academic Research"). Additionally, In order for both data sources to be congruent, i.e., refer to industry destinations rather than specific experiences, text referring to job roles (e.g., Quantum Physicist or Quantum Engineer) was also not included, resulting in 42 unique words categorised in the final dataset of **D2**. In this manner, both **D1** and **D2** referred only to industries.

It is important to note that each occurrence has been studied in the context of its text and classified into a single career destination, even if it appears multiple times within a single document. For instance, if the word “computing” appeared three times in a single document, it was only classified into one career destination.

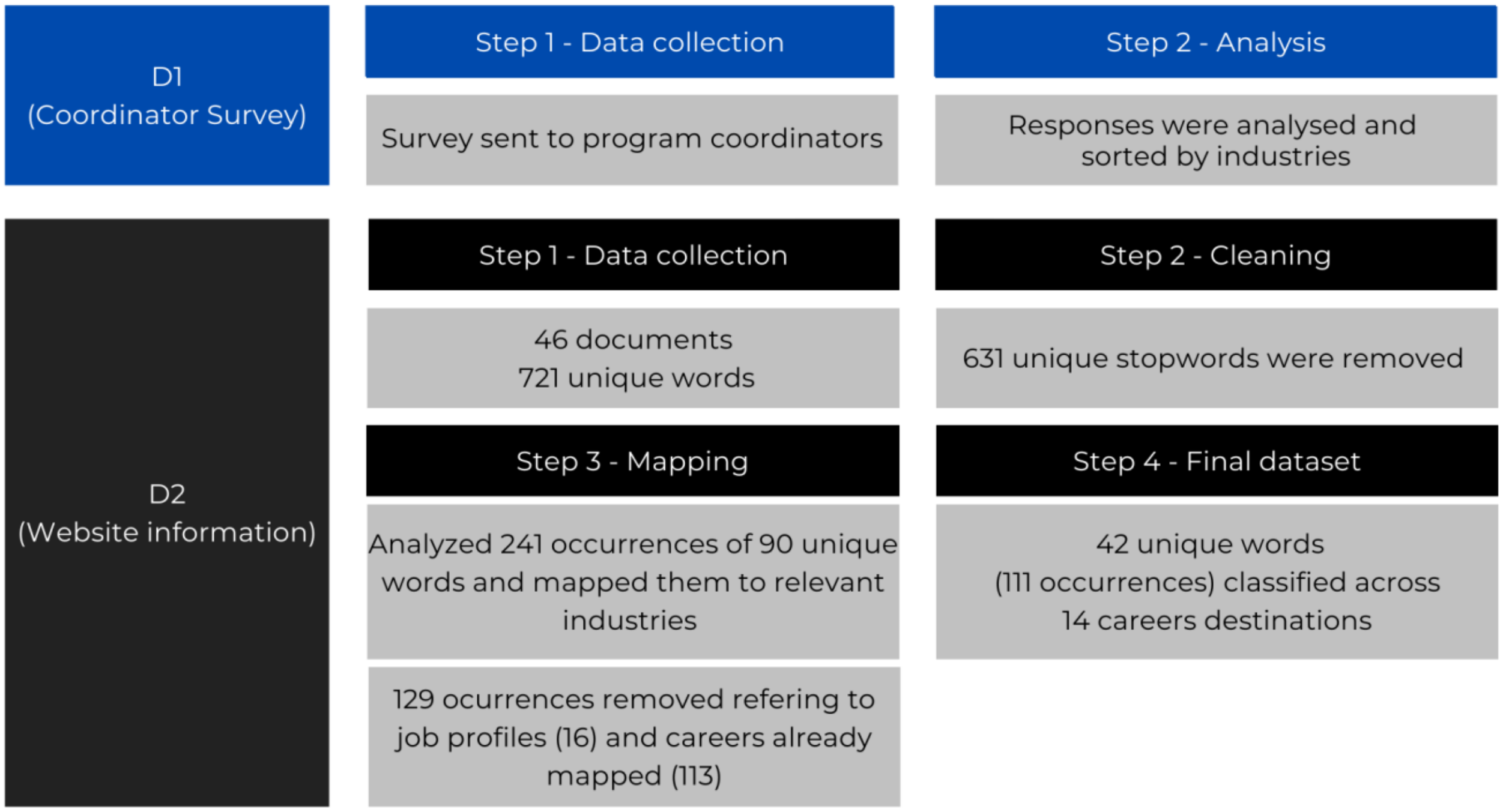

***Figure 1**. Methodology used for assigning graduate career outcomes obtained through program coordinators (D1) and masters web pages (D2).*

After an in-depth analysis, our final dataset consists in a final list of 42 unique words and a total of 111 occurrences have been classified across 14 distinct career destinations.

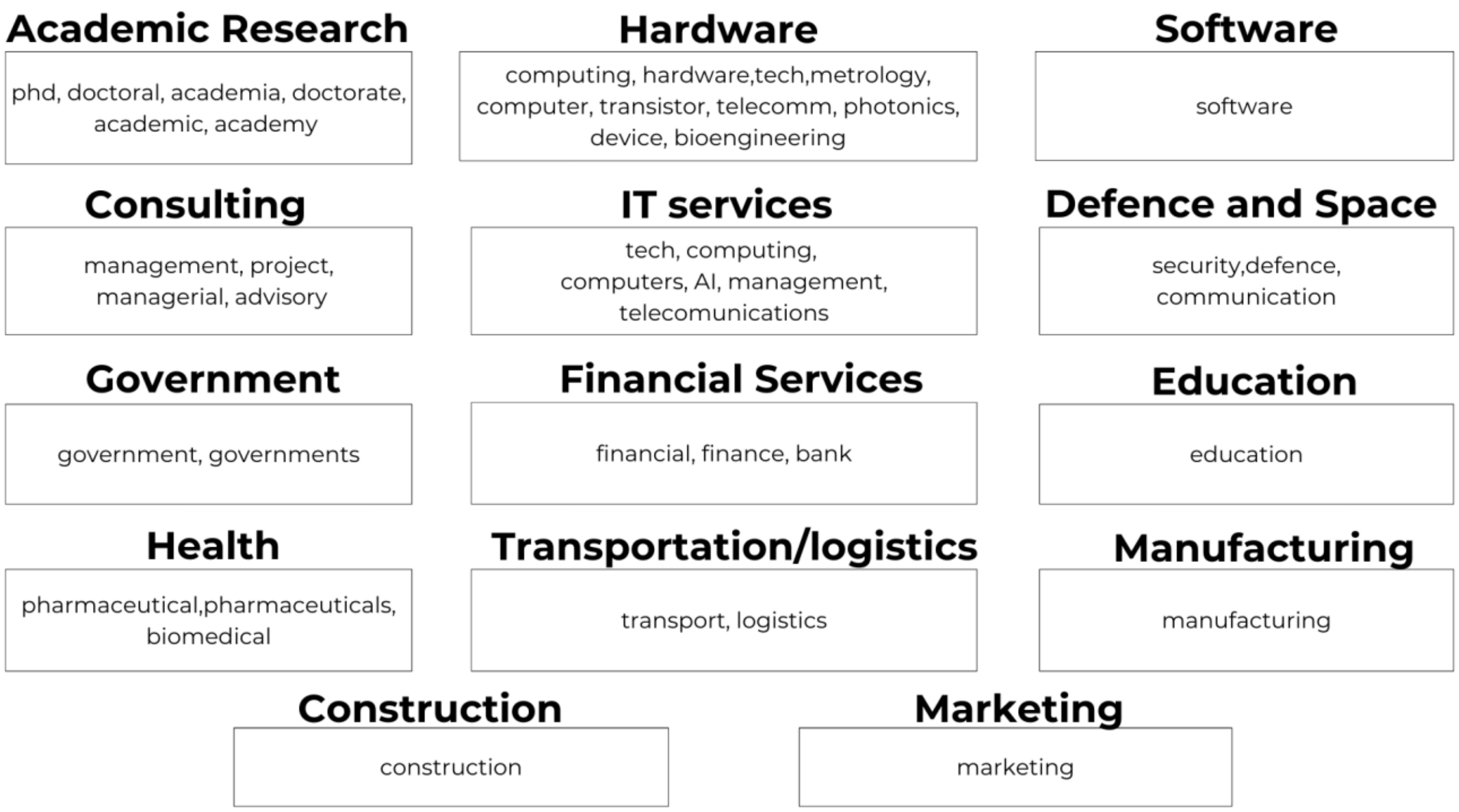

***Figure 2**. Unique words describing student career outcomes classified by industry, in order of greatest to least prevalence within the industry category. Note that each word was analysed in the complete context of the document, not as individual words.*

The final result of our analysis **(D2),** alongside the career information obtained from the program coordinators **(D1)** was merged in order to generate a final projection of possible career outcomes for students in the form of 14 industry destinations.

# 3. Results

86 Quantum technology master's programs worldwide were gathered, including primary master's (**67)**, and secondary master's **(19).**

## 3.1 Emergence of master's programs over time

Although there are specific master's programs on quantum technologies either in quantum computing, quantum sensing, quantum communication, quantum simulation, or engineering devices, there is also an evident progressive specialization in quantum technology of more general master's in physics, sciences or computing sciences, a trend we shall discuss further in the discussion section. Figure 3 illustrates the worldwide establishment of master's programs over time.

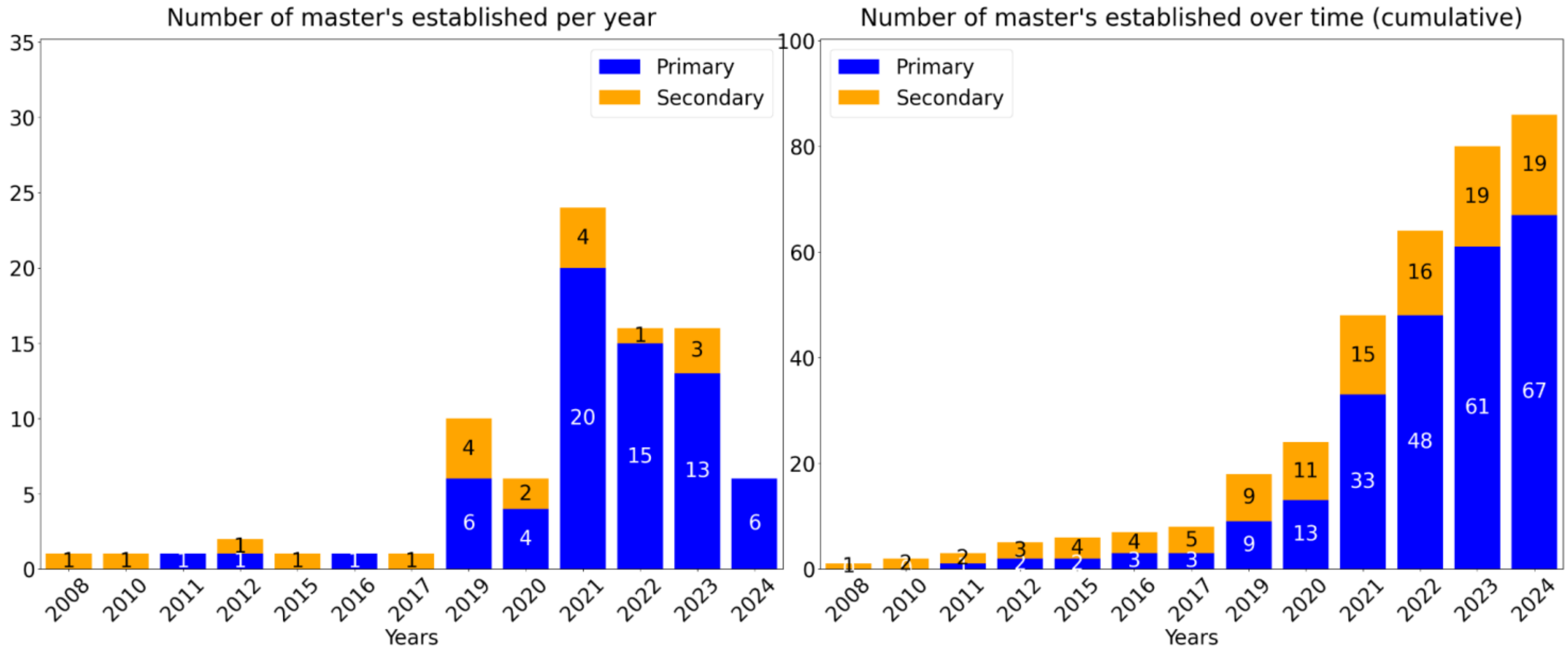

***Figure 3***.*Master's programs established worldwide over time, from 2008 to 2024, with a total of 67 primary master's and 19 secondary master's.*

The emergence of numerous master's programs in recent years suggests a growing trend. Most master's programs started after 2019, which is in line with the propagation of quantum strategies in many nations. Many countries cite[2],[12],[13],[14],[15],[17],[18],[19],[20],[21],[22],[23],[24],[25], have highlighted examples of master's programs or outlined the need for the creation of new master’s programs as part of these strategies. Given the several year time lag between publication of a national strategy and the successful launch of new programs, this may explain the substantial rise in the number of programs beyond 2021. Previous research recognised 40 universities with master's programs worldwide[2] as of 2022. Since then, it is clear that there has been a substantial scaling up worldwide. When referring to Figure 3, we can see that most of the growth in master’s programs began after 2020. In 2021, 24 of the currently running master’s programs were established (27.9%), whilst in each of 2022 and 2023, 16 master’s programs were established (18.6%). A total of 72.1% of all the currently running master’s programs began since the start of 2021, despite only representing a three year timeframe. This shows the magnitude of the growth of master’s programs around the world in the last few years.

## 3.2 Geographical distribution of master’s programs

As of now there are 86 master’s programs identified. Their geographical distribution is shown in Figure 4.

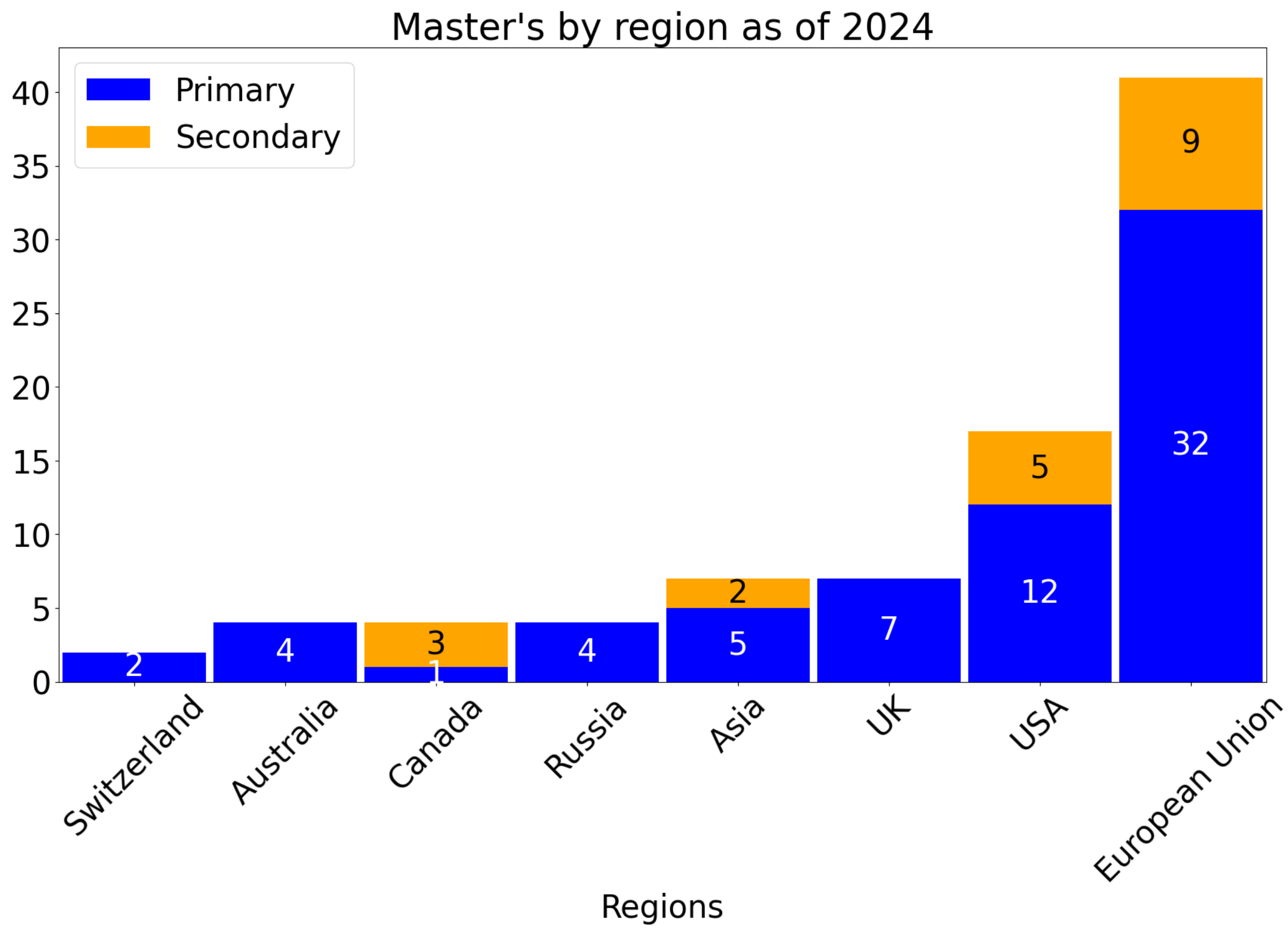

***Figure 4****. The establishment of Master's degrees worldwide per region. The figure shows a total of 86 masters across different regions.*

As of now, In Europe alone a total of 41 master's have been identified. The countries represented include Czech Republic (1), Denmark (2), France (9), Germany (9), Greece (1), Ireland (1), Italy (5), Netherlands (2), Poland (1), Spain (8), and Sweden (2). We note here that the countries with the most master's programs are France and Germany (9 programs each), in line with the largest national investment in QT. Spain, despite having limited QT research and development funding[39] hosts 8 master's programs and may therefore be well positioned with a workforce for future national quantum efforts.

As for Asia, the countries represented are Israel (2), Iran (1), Saudi Arabia (1), India (1), Thailand (1), and South Korea (1). We note that India [40],[41], Israel[15], Thailand[42], and South Korea[16] have Quantum National Strategies and in particular, the latter three emphasise the need to create or extend master's programs within their Quantum National Strategy.

Although there has been extensive research and funding in the field of QT in China[43], it is unclear whether there are any QT master's programs available in China. The lack of clear information regarding China has been noted previously by Ezratty[44], when referring to the Chinese government's investment in quantum technology. This should be borne in mind as a limitation of this study. For Non-EU, two countries are represented, Switzerland (2) and Russia (4). Finally, 32 master's degrees are distributed among the United States (17), United Kingdom (7), Australia (4) and Canada (4).

## 3.3 Faculties organising master's programs

We have clustered the faculties organising the master's programs into four main categories: Physics, Engineering, Computer Science and Joint Faculties. Whilst many programs are organised by faculties of Physics, since 2019 there is an increasing proportion of programs running from Joint Faculties, reflective of the increasing interdisciplinarity of QT. In general, this shows a commitment encompassing various learning approaches and perspectives. The diversity of competencies and skills required to meet the needs of the quantum industry would suggest there may be further continuous diversity of faculties working together to deliver competitive programmes to overcome the quantum talent shortage[3].

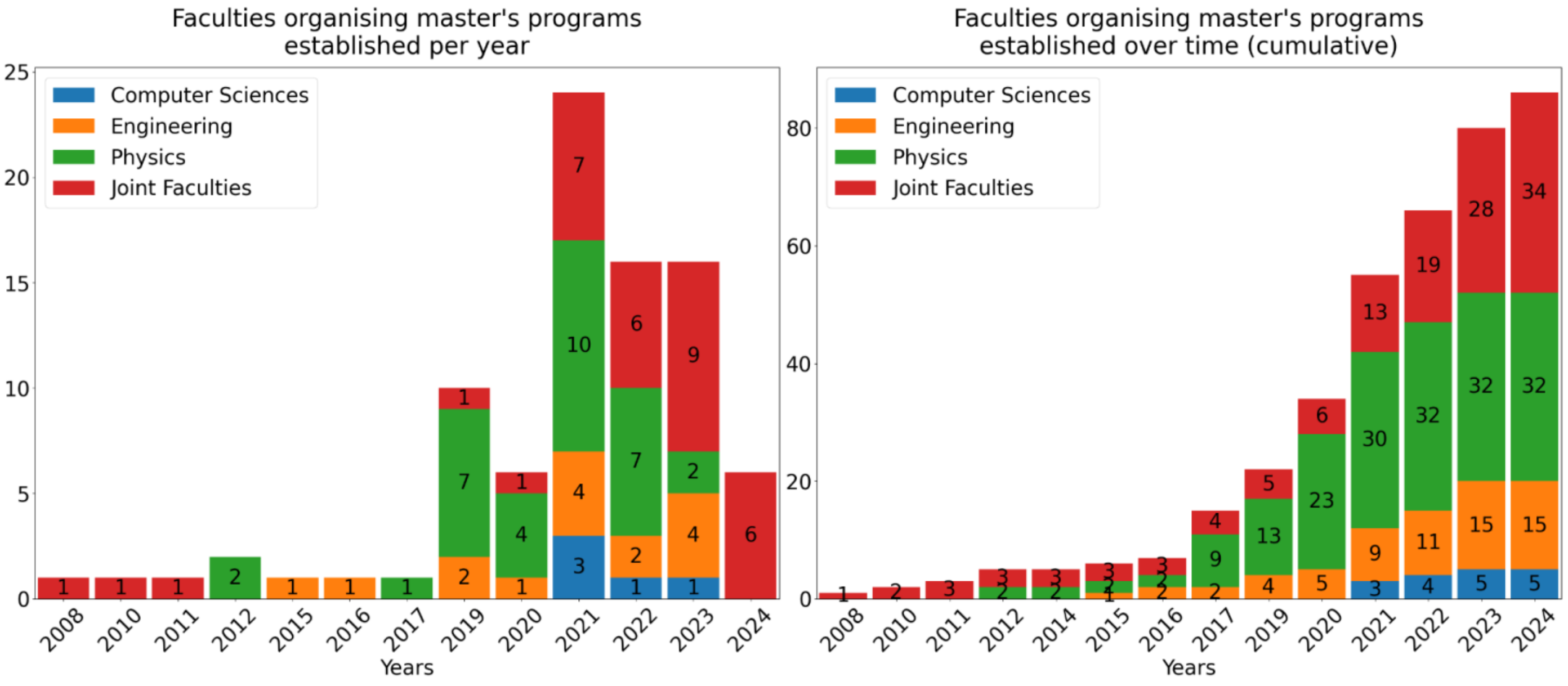

***Figure 5****. Faculties organising master's programs established per year, showing an increasing proportion of programs running from Joint Faculties.*

Reflecting on this journey towards interdisciplinary can present challenges on what concepts and courses should be taught and why those may be more relevant than others, depending on the tradition and relevance for the field as well as the diverse backgrounds of instructors[45,46]. As such, proposed recommendations such as "gathering subject matter experts" or "hiring a faculty cluster" could potentially facilitate going beyond classical physics departments towards new faculty lines[8]. In Europe, we have identified at least 10[47,48,49,50,51,52,53,54,55,56] jointly developed quantum master's which address elements such as the mobility of students for conducting semesters or doing research practices abroad. These master's are composed of several faculties which could mitigate some of the challenges associated with developing new master's programs, such as lacking existing research or faculty expertise in QT[29].

## 3.4 Provision of hands-on experience in master's programs

The information obtained regarding the internship offerings was filtered by analysing the curriculums and websites of each master's program and by gathering information from program coordinators. Looking into the internship offering of each program, we have identified four main categories: "internship industry", "internship university" and internships offered either in the university or industrial settings.

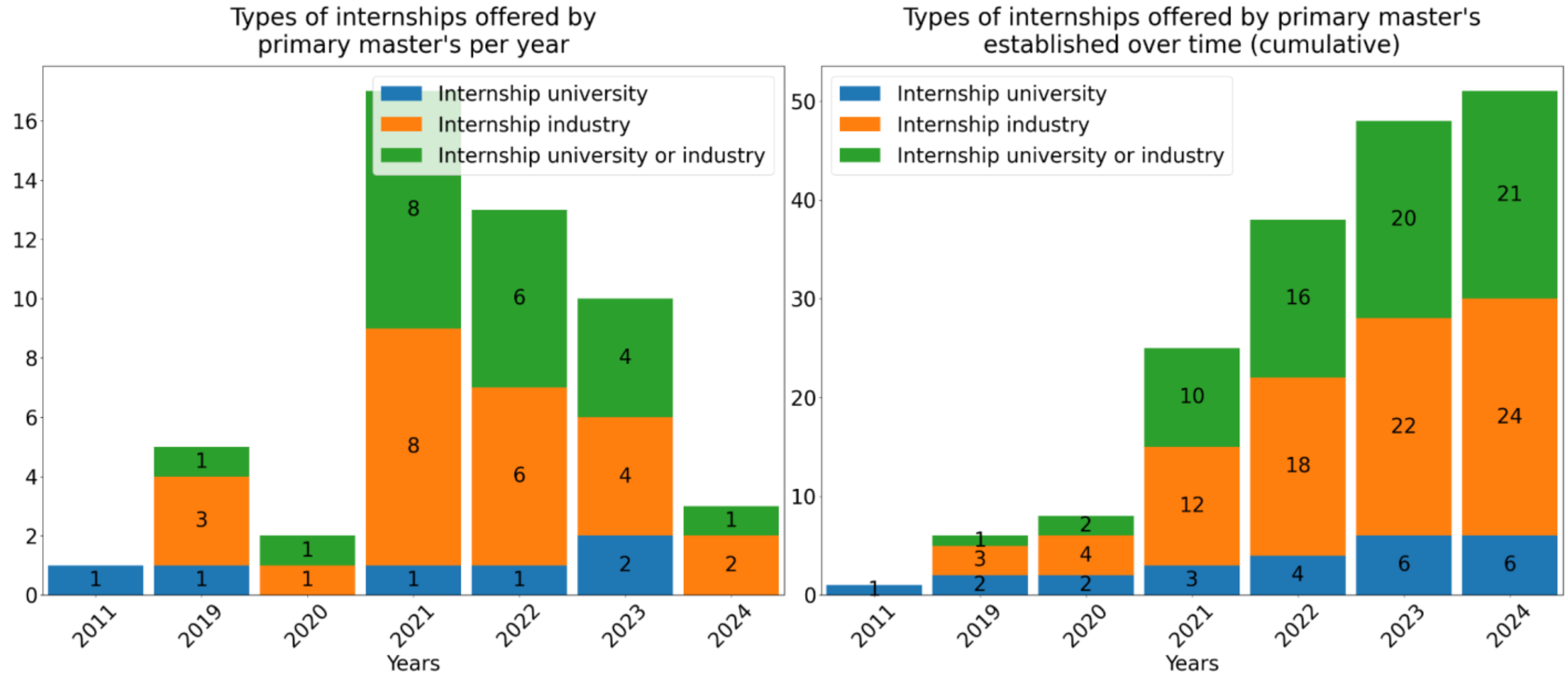

***Figure 6**. Type of internships offered by Primary Master's established over time. The figure shows the Increasing trend towards internships carried out in an industry setting.*

The structure of the internship may differ from one university to another, ranging from a several month on-site commitment at a company, to a weekly visit to a research lab for collaboration on a specific project. We considered all cases of provision of hands-on experience for ther purpose of this category, with the source categorised by university, company, or unspecified/either.

It is clear from our findings that there is an increasing trend towards internships carried out in an industry setting, which is in line with the increasing commercialisation of QT[57], and may be a valuable means to connect graduates to industry directly. Prior to 2019, none of the established programs offered internships in companies, whilst as of 2020 greater than 50% of newly launched program's offer an industry internship.

## 3.5 Career outcomes for master's graduates

We have described career outcomes by the industries in which graduates are expected to enter, as reported by program coordinators. It is important to acknowledge that while many of the programs are at a stage of relative infancy, they may not have reliable data following up on their graduates, and therefore this information is not necessarily based on follow-up studies conducted with graduates. Rather, it may be an estimation based on the content of the program. Furthermore, we note that while our data may inform us of the potential industry sectors master's graduates will go to, additional research would be needed to understand the specific roles and tasks that will be required within these sectors[4,5,6].

***Figure 7.*** *Career outcomes bar chart, primary and secondary masters combined, worldwide.*

One question that arises from previous findings [2,3,8] is whether quantum master's programs are designed to solve the industry's needs by supplying graduates directly to fill job roles, or whether they are stepping stones towards doctoral studies. Whilst our findings do not provide numerical information on the career destinations of QT graduates, they do suggest that master's degrees can lead to jobs in many industries, not only to further studies. Only 17.6% of the career outcomes indicated by program websites and coordinators refer to academic career paths, while 82.3% refer to industry careers. So at least from the perspective of program organisers, careers in industry are certainly intended for master's graduates.

The studies may help students to gain specialist experience in QT[30], which they may bring to satisfy demands in different emerging areas by providing their expertise in sectors such as IT, hardware, software or governmental institutions. The "quantum-adjacent" industries[5] will benefit from having graduate master's students who could significantly serve as an essential chain within the company, as they are expected to gather significant theoretical knowledge and practice skills during their master's programs.[1,2,8] Graduate students could contribute to the expansion and further realisation of the field, thus overcoming the "skill-gap" present in the emerging industry[7]. One of the possibilities for attracting talent to these companies is early collaboration between the departments in charge of the master's program either through internships, conferences or other ways to meet the demand for these professionals and offer an "alternative pathway" into the quantum industry that does not involve obtaining a PhD[1,2,8].

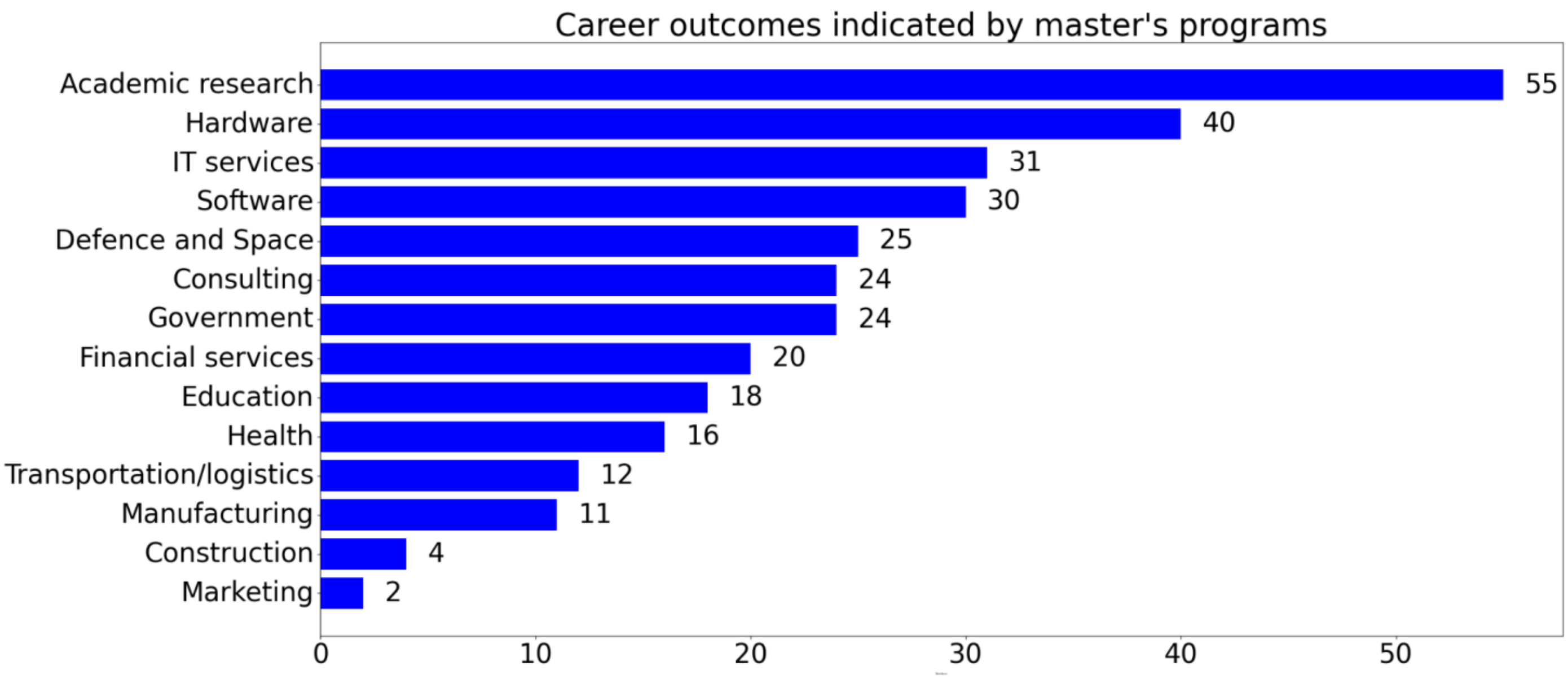

## 4. Discussion

We have observed a substantial growth in the number of QT master's programs, both primary and secondary, over the last decade. Master's students will access the field through either primary or secondary master's. The increasing trend towards augmentation courses and secondary master's shows that QT is starting to spin out further, and there is a need to allow students from non-physics backgrounds access to quantum knowledge.

## 4.1 Increasing interdisciplinary in QT Education

One clear trend we have seen is the increasing interdisciplinary nature of the applications of QT. As we no longer need only physicists, but also engineers, medics, business people, and graduates with a diverse range of backgrounds. This is reflected in the significant increase in the number of Joint Faculties since 2019, when most of the programs have been established. The number of programs run by Joint Faculties has almost doubled from 2019, from representing 22.7% of all programs established up to and including 2019, to 39,5% of those existing up to and including 2024, as reflected in Figure 8. We also note that the first programs run by Computer Science faculties began only in 2021, reflecting the increasing realisation of Quantum Computing since claims of Quantum advantage in 2019[58]. Over time since then, it has become clear that computer science graduates may benefit significantly from QT experience[59].

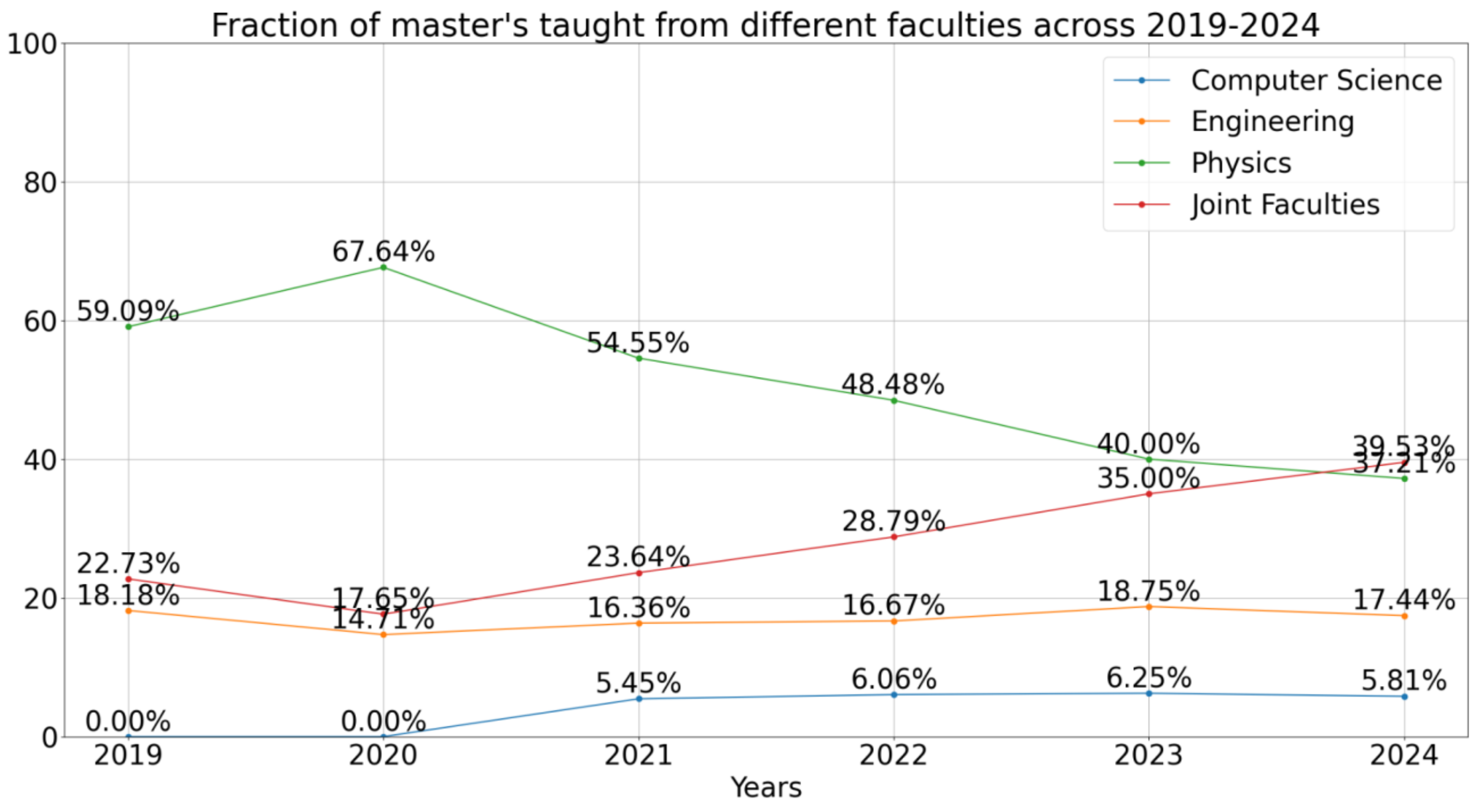

***Figure 8**. Fraction of Masters taught from different faculties across 2019-2024*

## 4.2 Joint master programs as a national strategy

Within Europe, we have identified 10[47],[48],[49],[50],[51],[52],[53],[54],[55],[56] jointly developed quantum master's, while no such joint programs exist outside of Europe. These joint degrees have been in the scope of the European political agenda since 2001 to strengthen the educational landscape[60] and promote interdisciplinarity[61]. They can also help to address program-enriching elements such as the mobility of students for conducting semesters or doing research practices abroad, enhancing the career opportunities of these students by expanding their network and providing a better understanding of the job market. They may also go some way towards fulfilling the objective signed by twenty-three member states of the EU, Europe's Quantum Declaration[62], taking coordinated action to implement the "training measures necessary to support and grow the EU quantum ecosystem".

Outside of Europe, nations such as South Korea[16], or South Africa[63], address the need to create and share a common curriculum in QT among diverse faculties in their national Quantum strategies, aiming to merge disciplines such as physics, engineering, and computer science to boost a multifaceted quantum technology education. South Africa specifically refer to the creation of interdisciplinary Master programs[63] . This collaborative approach could benefit from the optimization of resources by sharing curricula, expertise and educational materials, bringing together more students from different backgrounds and therefore increasing accessibility to QT, and supporting a more diverse workforce.

It seems likely that the European education landscape, and the history of the Bologna Process[64], filters down to the Quantum Technology educational landscape, resulting in many more joint programs than other nations. Elsewhere, set up of joint programs may be more difficult, and a more flexible approach may be needed, for example organising international exchanges through the global entanglement exchange[65].

## 4.3 Quantum master's can be a valuable boon to the industry

Previous research has questioned the value of the master's programs in meeting the needs of the workforce. Hughes et al.[1] noted that "If a PhD is needed for the highly specific quantum jobs, then this places doubt on how many new highly-focused quantum master's programs are needed for students wanting to enter the quantum workforce". However, it may be that this perspective has shifted somewhat rapidly in the past several years. Even as of the Hughes et al's survey in 2020, many quantum-specific jobs outside of academia were fairly accessible to master's graduates. For example roughly 50% of their surveyed companies suggested that master's graduates could be suitable for the role of "Quantum Algorithm Engineer"[1].

In the last few years, many more master's programs are now in interdisciplinary fields (as shown in Figure 8), providing students careers in a wide variety of industries, and providing quantum-specific expertise to students from areas such as Computer Science and Engineering. We would expect that these graduates, with a wider variety of skills, would be well suited to both quantum-specific and more generalist technical roles, and so the proportion of job roles which can be fulfilled by master's graduates may well increase.

Furthermore, with the number of master's programs so significantly increasing, we anticipate that many more companies may consider hiring their graduates directly, since they are more aware of the learning outcomes of these students and the QT master's program as a possible route to quantum expertise. Future research done in the European Quantum Readiness Center[36] will investigate this quantitatively, using data from job posts worldwide. In their review of the quantum workforce landscape, Kaur and Venegas-Gomez[2] noted that master's programs may not be sufficient to meet the needs of the workforce, due to "education costs, degree value, and accessibility"[2]. This is another area which appears to be changing rapidly, as the increasing number of master's programs, their interdisciplinarity, and worldwide presence will mitigate many of these issues.

## 4.4 National efforts to enhance Quantum Technology programs

Aside from developing new master's, another approach to providing QT skills to graduates is to enhance existing programs with QT experience. There are an increasing number of these being developed across the levels of Higher Education, including among bachelor's, master's, and PhD programs. At the bachelor's level, minors[66],[67],[68],[69],[70] can be offered to graduate students from many STEM disciplines with fewer credits than a master's program, but providing some quantum experience, and perhaps more importantly, interest and inspiration in the topic.

We have identified a trend in national funding which makes use of this approach, which we term “Quantum program enhancements”, to enable projects which are designed chiefly to develop content that is then used to augment many different study programs with additional quantum experience. This is in contrast to the joint master’s programs, which develop their own uniquely tailored program. For example, in Europe, the project *DigiQ: Digitally Enhanced Quantum Technology Master* is developing a set of shared QT courses which are used by many different master’s programs (a total of 16), including a mixture of newly developed and previously existing, enhanced, primary and secondary master’s programs[71].

Such a large upscaling of European QT master’s would only be possible with an efficient solution like DigiQ. While joint programs such as those organised in the Erasmus Mundus framework[72] can be effective, they are limited in feasibility by costs and national accreditation procedures[73]. As such, these program enhancements can be a far more efficient solution to offer more interdisciplinary, hands-on QT education, while minimising expenses associated with setting up new programs entirely from scratch.

Many nations have begun funding such community-led projects (see Fig 10), pathfinding approaches to developing enhancements for bachelor’s, master’s, and PhD programs[71],[77],[78],[79],[81],[82],[83],[85],[86]. In general, these initiatives are nationally funded and involve the development of educational programs, courses, internships, and collaborative projects aimed at attracting, producing and sharing educational resources from bachelor's to Phd students. Common features include strengthening collaboration between academia and industry through internships, as well as a focus on building a strong national quantum ecosystem addressing the growing demand for skilled professionals in this field. For interested readers, additional information is available about each of these national projects in Appendix 2.

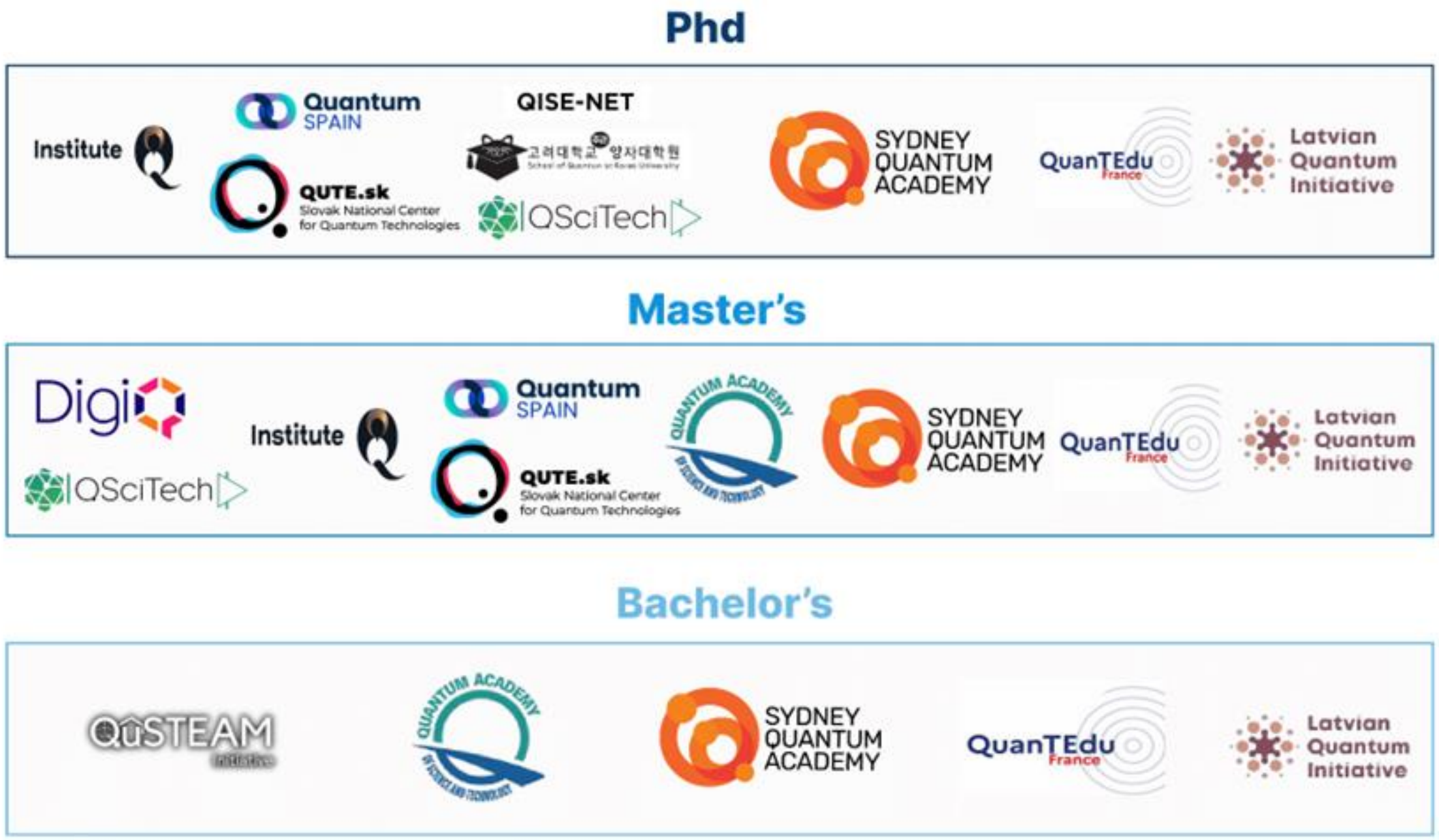

***Figure 9****. National projects in Quantum Technology Education worldwide*

## 5. Conclusion

Given the increasing number of master's worldwide, and their prevalence within the Quantum National Strategies of many nations, our findings suggest that master's programs may become more than simply a stepping stone to a PhD. Rather they can be a standalone qualification which shortcuts the long timeline associated with doctoral studies, to offer direct access to the quantum industry. The increasingly interdisciplinary opportunities, and many more internships placing students directly working with companies, can help to boost student opportunities to where more and more jobs are recruiting from this increasing pool of graduates. In turn, we anticipate this may help to grow the quantum industry in many nations.

For the purpose of this paper, we have analysed the career prospects of master's graduates from the perspective of the program websites and program coordinators. Our findings may add to the growing body of research exploring the needs of the quantum industry, which to date has primarily comprised information from the perspective of employers. It is valuable to also have the perspective of program coordinators which we have provided. However there is still a distinct lack of a student voice in this area. To have a more complete understanding of what role universities can and should take in order to supply the needs of the quantum industry, it is essential to take the supply perspective into account, such as through interview studies with students exploring their career outcomes after graduating from master's studies. Given the quantum master's as a fruitful option for careers in industry, it will also become increasingly important to change the perception of undergraduates and school students that quantum science is only for "geniuses" or requires years of dedicated doctoral study[11]. Outreach has an essential role to play here in demystifying the field[74], and with 2025 dedicated by UNESCO as "International Year of Quantum Science and Technology", we may well see an acceleration of these efforts[75].

## List of abbreviations

ARTEQ - Année de Recherche en Technologies Quantiques

CS - Computer Science

DigiQ - Digitally Enhanced Quantum Technology Master

ECE - Electrical and Computer Engineering

EU - European Union

InstituteQ - The Finnish Quantum Institute

IT - Information Technology

QDOC - Quantum Doctoral Pilot Programme

QISE - Quantum Information Science and Engineering

QISE-NET - Quantum Information Science and Engineering Network

QIST - Quantum Information Science and Technology

QSciTech - Bridging the Gap between Quantum Science and Quantum Technologies

QT - Quantum Technology

QTEdu - Quantum Technology Education

QuSTEAM - Convergent Undergraduate Education in Quantum Science, Technology, Engineering, Arts and Mathematics

Qute.sk - National Center for Quantum Technologies

SQA - Sydney Quantum Academy

STEM - Science Technology Engineering Mathematics

UASs - Universities of applied sciences

UK - United Kingdom

UNESCO - United Nations Educational, Scientific and Cultural Organization

USA - United States of America

# Declarations

## Availability of data and materials

The majority of data sources used during the preparation of this manuscript is available in the Appendices. Additional data used for analysis is available as the supporting information, excepting data obtained through email survey, which is provided in aggregate for privacy protection.

## Competing interests

The authors declare no competing interests.

## Funding

The authors acknowledge funding from the Digital Europe Programme for the project DigiQ: Digitally Enhanced Quantum Technology Master (DigiQ) under grant agreement ID 101084035. The authors acknowledge funding from Horizon Europe for the project Quantum Flagship Coordination Action and Support (QUCATS) under grant agreement ID 101070193.

## Authors' contributions

J.S initiated the project from which the research came. S.G and B.M. designed and conducted the research. S.G and B.M. primarily wrote the paper. All authors reviewed the manuscript.

## Acknowledgements

## Authors' information

Borja Muñoz is a researcher at the Center for Hybrid Intelligence at Aarhus University. He focuses on the quantum technology ecosystem exploring it from a societal approach.

Simon Goorney is a researcher in the European Quantum Readiness Center, based at Aarhus University. He is responsible for implementation of education programs for Quantum Technology, including Europe’s largest QT education project DigiQ. He also studies the changes in the education and industry landscapes which arise from community innovations.

Jacob Sherson is Professor of Management in Aarhus University and of Physics at the Niels Bohr Institute, Copenhagen University. He is director of the Center for Hybrid Intelligence and the European Quantum Readiness Center, and a leading figure in educational reform for Quantum Technologies in Europe.

# Appendix 1 - List of masters worldwide

**Table 1** List of masters worldwide

| Number | Country | City | University | Name of the program | Primary or secondary |
|---|---|---|---|---|---|
| 1 | Australia | Brisbane | University of Queensland | Master of Quantum Technology | P |
| 2 | Australia | Canberra | The Australian National University | Master of Science in Quantum Technology | P |
| 3 | Australia | Canberra | The Australian National University | Master of Science (Advanced) in Quantum Technology | P |
| 4 | Australia | Melbourne | La Trobe University | Master of Quantum Information Technology | P |
| 5 | Canada | Calgary | University of Calgary | Master of Quantum Computing | P |
| 6 | Canada | Toronto | University of Toronto | Quantum Computing Concentration | S |
| 7 | Canada | Waterloo | University of Waterloo | Master Quantum Information Program | S |
| 8 | Canada | Waterloo | University of Waterloo | Msc Physics - Quantum Technology specialization | S |
| 9 | Czech Republic | Prague | Czech Technical University (CTU) | Quantum Technologies | P |
| 10 | Denmark | Copenhagen | University of Copenhagen (KU) & Technical University of Denmark (DTU) | Master of Science (MSc) in Quantum Information Science | P |
| 11 | Denmark | Copenhagen | Technical University of Denmark (DTU) | Master of Science (Msc) in Engineering Physics - Study line in Quantum Engineering | S |
| 12 | France | Dijon | University Bourgogne Franche-Comté (UBFC), Aarhus University (AU) & The University of Kaiserslautern-Landau (RTPU) | Erasmus Mundus Master Quantum Technologies and Engineering (QuanTEEM) | P |
| 13 | France | Paris | Sorbonne University | Master of Computer Science - Quantum Information (QI) | P |
| 14 | France | Grenoble | Université Grenoble Alpes | Master Quantum Information Quantum Engineering | P |
| 15 | France | Paris | Université de Paris, the Ecole Polytechnique, Ecole d'Ingénieur Denis Diderot (EIDD), and Politecnico of Turin | Master Quantum Devices | S |
| 16 | France | Paris | Université Paris-Saclay | Master Program Computer Science (M1 / M2 in Quantum and | S |
| | | | | Distributed Computer Science) | |
| 17 | France | Paris | Université PSL (Paris Sciences & Lettres) | Master Quantum Engineering | P |
| 18 | France | Paris | Universite Paris-Saclay, Sapienza Universitá de Roma, University of Porto & University of Toronto | Quantum Research Master Education Network – Erasmus Mundus QUARMEN | P |
| 19 | France | Strasbourg | University of Strasbourg | Quantum technologies, European program | S |
| 20 | France | Paris | l'École normale supérieure - PSL / Sorbonne Université, Université de Paris, Université Paris-Saclay and Institut Polytechnique de Paris | Quantum physics: From the foundations to quantum technologies | S |
| 21 | Germany | Saarbrücken | Saarland University | Master's Program Quantum Engineering | P |
| 22 | Germany | Aachen | RWTH Aachen University | Master study-track on Quantum Technologies | S |
| 23 | Germany | Munich | Technical University of Munich (TUM) & Ludwig Maximilian | Master of Science (M.Sc.)Quantum Science & Technology | P |
| | | | University of Munich (LUM) | | |
| 24 | Germany | Würzburg | University of Würzburg | Master's Degree Program (M.Sc.) Quantum Technology | P |
| 25 | Germany | Deggendorf | Deggendorf Institute of Technology | High Performance Computing / Quantum Computing MSc | P |
| 26 | Germany | Hannover | Leibniz University Hannover | Quantum Engineering (Master of Science) | P |
| 27 | Germany | Tübingen | University of Tübingen | Msc Advanced Quantum Physics | P |
| 28 | Germany | Braunschweig | Technische Universität Braunschweig | Quantum Technologies in Electrical and | P |
| | | | | Computer Engineering (QTEC) | |
| 29 | Germany | Kaiserslautern | University of Kaiserslautern-Landau (RTPU) | Master of Science Quantum Technologies | P |
| 30 | Greece | Xánthi | Democritus University of Thrace (DECE) | International M.Sc. Program in Quantum Computing and Quantum Technologies | P |
| 31 | India | Bangalore | Indian Institute of Science | M Tech.Program in Quantum Technology | P |
| 32 | Iran | Teheran | Pasargad Institute for Advanced Innovative Solutions | International master program "Quantum Materials, Energies, and Technologies (QMET) | P |
| 33 | Ireland | Dublin | Trinity College Dublin | MSc Quantum Science and Technology | P |
| 34 | Israel | Jerusalem | The hebrew University of Jerusalem | Tracks in Quantum Information Science and Technology | S |
| 35 | Israel | Haifa | Technion Israel Institute of Technology | Master's degree with a specialization in Quantum Science & Technology | S |
| 36 | Italy | Pavia | University of Pavia | MSc in Physical Sciences, Physics of Quantum Technologies | S |
| 37 | Italy | Naples | University of Naples Federico II | Master's Degree in Quantum Science and Engineering | P |
| 38 | Italy | Torino | Politecnico di Torino | Master's Degree Program  Quantum Engineering | P |
| 39 | Italy | Venice | University of Venice | Executive Master in Quantum Machine Learning | P |
| 40 | Italy | Trieste | University of Trieste | Master’s Degree program in Scientific and Data-Intensive Computing with quantum computing concentration | S |
| 41 | Netherlands | Delft | Delft University of Technology (TU Delft) | MSc  Quantum Information Science & Technology | P |
| 42 | Netherlands | Amsterdam | University of Amsterdam | Master in Quantum Computer Science | P |
| 43 | Poland | Gdańsk | University of Gdańsk | Quantum Information Technology | P |
| 44 | Russia | Moscow | MSU Quantum Technology Centre | Master's Program Quantum Computing | P |
| 45 | Russia | Moscow | MSU Quantum Technology Centre | Master's Program Applied Quantum Communication | P |
| 46 | Russia | Moscow | MSU Quantum Technology Centre | Master's Program Quantum and Optical Technologies | P |
| 47 | Russia | Novosibirsk | Novosibirsk State University | Quantum Technologies and Cryptography | P |
| 48 | Saudi Arabia | Dhahran | King Fahd University of Petroleum and Minerals | Professional Master in Quantum Information and Computing | P |
| 49 | South Korea | Seoul | University of Korea | Quantum Information Science Convergence Major | P |
| 50 | Spain | A Coruña | University of A Coruña, University of Santiago de Compostela and University of Vigo | Master's Degree in Quantum Information Science and Technology | P |
| 51 | Spain | Leioa | The University of the Basque Country (UPV/EHU) | Master in Quantum Science and Technology | P |
| 52 | Spain | Barcelona | University of Barcelona (UB), Universitat Autònoma de Barcelona (UAB), Universitat Politècnica de Catalunya (UPC), The Institute of Photonics Sciences (ICFO), Barcelona Supercomputing Center (BSC-CNS), Instituto de Física de Altas Energías (IFAE), Catalan Institute of Nanoscience & Nanotechnology (ICN2) | Master in Quantum Science and Technology | P |
| 53 | Spain | Madrid | University Carlos III of Madrid (UC3M) | Master in Quantum Technologies and Engineering | P |
| 54 | Spain | Madrid | Technical University of Madrid (UPM) | Master in Quantum Computing Technology | P |
| 55 | Spain | Madrid | Nebrija University | Master´s Degree in Quantum Computing | P |
| 56 | Spain | Madrid | Menéndez Pelayo International University (UIMP) & Spanish National Research Council (CSIC) | Master's Degree in Quantum Technologies | P |
| 57 | Spain | Logroño | International University of La Rioja (UNIR) | Master’s Degree in Quantum Computing | P |
| 58 | Sweden | Stockholm | KTH Royal Institute of Technology | Msc Engineering Physics Track in Quantum Technology | S |
| 59 | Sweden | Uppsala | Uppsala University | Master's Programme in Quantum Technology | P |
| 60 | Switzerland | Lausanne | École polytechnique fédérale de Lausanne (EPFL) | Quantum Science and Engineering | P |
| 61 | Switzerland | Zürich | ETH Zürich | Master Quantum Engineering | P |
| 62 | Thailand | Chiang Mai | Chiang Mai University | Master of Science Program in Quantum Science and Technology | P |
| 63 | United Kingdom | London | University College London (UCL) | Quantum Technologies MSc | P |
| 64 | United Kingdom | Sussex | University of Sussex | Quantum Technology MSc | P |
| 65 | United Kingdom | Sussex | University of Sussex | Quantum Technology Applications and Management MSc (online) | P |
| 66 | United Kingdom | Bristol | University of Bristol | MSc Optoelectronic and Quantum Technologies | P |
| 67 | United Kingdom | Glasgow | University of Strathclyde | Msc Quantum Technologies | P |
| 68 | United Kingdom | Glasgow | University of Glasgow | MSc Quantum Technology | P |
| 69 | United Kingdom | Guildford | University of Surrey | Applied Quantum Computing - Msc | P |
| 70 | United States | Los Angeles | University of California Los Angeles (UCLA) | Master Of Quantum Science And Technology | P |
| 71 | United States | Bloomington | Indiana University Bloomington | MSc in Quantum Information Science | P |
| 72 | United States | Golden | Colorado School of Mines | Quantum Engineering Graduate program | P |
| 73 | United States | Newark | University of Delaware | MS Quantum Science and Engineering | P |
| 74 | United States | Los Angeles | University of Southern California, Viterbi | Master of Science in Quantum Information Science | P |
| | | | School of Engineering | | |
| 75 | United States | Madison | University of Wisconsin-Madison | MS in Physics-Quantum Computing | S |
| 76 | United States | Buffalo | University of Buffalo | Engineering Science (Quantum Science and Nanotechnology) | S |
| 77 | United States | Durham | Duke University | Study Track Quantum Software/Hardware | S |
| 78 | United States | Kingston | The Univeristy of Rhode Island | MS in Quantum Computing | P |
| 79 | United States | Hoboken | Stevens Institute of Technology | Master of Science in Quantum Engineering | P |
| 80 | United States | Tucson | The University of Arizona | M.S. in Optical Sciences: Quantum Information | P |
| 81 | United States | New York City | Columbia University | M.S in Quantum Science and Technology | P |
| 82 | United States | Fairfax | George Manson University | MS in Physics with QISE Concentration | S |
| 83 | United States | San Jose | San José State University | MS in Quantum Technology | P |
| 84 | United States | Evanston | Northwestern University | specialization in Quantum Computing, Sensing & Communications | S |
| 85 | United States | Laurel | Capitol Technology University | Master of Research (MRes) in Quantum Computing | P |
| 86 | United States | New York | Stony Brook University | M. Sc. in Quantum Information Science and Technology | P |

# Appendix 2 - National Projects in Quantum Education

## Europe; DigiQ: Digitally Enhanced Quantum Technology Master

DigiQ[71]: DigiQ, supported by the EU's Digital Europe Programme, represents a tremendous effort to coordinate 16 European Master programs at the European Level. The project was born as a result of a previous experience with QTOM. DigiQ coordinated by the University of Aarhus aims to support the existing need for a quantum-talent workforce. The program is creating innovative modules and specialist courses (e-learning and on-site) structured under the Quantum Technology Competence Framework. Among its offerings, the 4 pan-European Quantum networks offer students a year-long social structure and communication platform with staff at universities and companies across Europe, including access to internships and QT events such as hackathons, workshops, and summer schools. The networks are thematic, covering topics such as science communication, quantum computing, concrete applications of quantum information science and hands-on experience in industrial settings.

**France: QuanTEDU**

Within the French National Strategy, the QuanTEdu project coordinated by the University of Grenoble Alpes works with 21 universities to provide quantum technology education from pre-university to doctoral levels. The large consortium also includes industrial partners to meet the demand for a diverse quantum workforce. The program is designed to meet the concrete actions established in the French National Strategy while ensuring a dynamic quantum ecosystem in France[79]. The program ARTEQ "Interdisciplinary Training Program in Quantum Technologies" is one clear example of a program aimed at master students from STEM disciplines oriented towards research and with the possibility of conducting research internships in a private or public laboratory[76].

## Canada: QSciTech

QSciTech. Graduate students (PhDs and MScs) in physics, engineering, computer science, mathematics and other disciplines are targeted by the project QSciTech for the purpose of expanding their knowledge and hands-on experience in quantum technology. Led by the University of Sherbrooke and funded by NSERC, a total of 7 universities form part of this consortium. Among the courses developed, which include Quantum Mechanics for non-physicist, Engineering Design & Project Management or Entrepreneurship and Scientific Research, the project also offers a paid internship with a duration of 4 to 6 months in a company[77].

## South Korea: School of Quantum at Korea University.

This program is led by the University of Korea with the goal of “developing a convergence-sharing university system for the intensive training of quantum information science research and industry experts”[81]. Nine universities are part of this consortium. Students are encouraged to select the desired university and advisor and apply for a major course in the field of QIS, which leads directly to a doctoral program. Internship opportunities are offered at domestic and overseas research institutions or companies through the program.

## Japan: QAcademy

QAcademy. Another example of an interdisciplinary program is the one led by the National Institute of Informatics in partnership with the University of Tokyo, Nagoya University, Kyushu University and Keio University. QAcademy developed a joint curriculum including online courses, internship programs as well and other learning resources for quantum technology. In a similar way to other initiatives mentioned, the program uses digitala enhancements, to expand access to specialist areas by sharing educational resources and courses across the partner universities[78].

### USA:QuSTEAM

QuSTEAM[79], composed by a large consortium of 23 partners and coordinated by Ohio State University offers quantum curricula and educational opportunities across disciplines. They are working on developing “campus-to-industry connections” including industry involvement and research practice. In addition, courses are targeted towards instructors, administrators and learners.

### USA:QISE-NET

QISE-NET[82], is a national training program targeted to graduate students in QIS. Co-lead by the University of Chicago and Harvard University and managed by the Chicago Quantum Exchange, it offers three years of funding for students and provides students mentorship to work on research questions as well as have internships and extended visits in company labs.

### Australia:Sydney Quantum Academy

Sydney Quantum Academy[80]. The SQA, which is backed by the NWS government, has partnered with some of the leading universities in Australia, including the University of Sydney, UNSW, UTS, and Macquarie University. They offer a diverse range of educational opportunities, from undergraduate to master's and PhD, along with scholarships, courses, and units of study. Likewise, the SQA collaborates with industry to provide internships to students, giving them valuable experience and exposure to future career opportunities.

### Spain:Quantum Spain

Quantum Spain[83] Within the national efforts to build a quantum computer, Quantum Spain is strengthening the educational system through TalentQ[84]. The program covers educational courses, seminar opportunities for Phd students and list a existing offer of quantum masters and phd offering based in Spain.

### Latvia:Latvian Quantum Initiative

Latvian Quantum Initiative[85]. 4 partners are involved as part of the Latvian Quantum Initiative, The University of Latvia (UL), Riga Technical University (RTU) and The Institute of Solid State Physics (UL), and the Institute of Mathematics and Computer Science (LU). The initiative focuses on research as well as education, where modules are being developed for students of different levels (bsc,msc and phd) as well as further education programmes. PhD and postdoctoral research opportunities are also offered on their website.

## Finland: InstituteQ

InstituteQ[86] The Institute offers the program Quantum Doctoral Pilot Programme (QDOC), financed by the Finnish Ministry of Education. Aalto University, University of Helsinki and the VTT Technical Research Center of Finland are part of this consortium . Shared curriculum is also offered through the University of Helsinki and Aalto University as part of their agreement, “Cooperation on quantum technology education in the Helsinki metropolitan area”, allowing undergraduate and graduate students to enrol in any of the courses offered.

## Slovakia: Qute.sk

Qute.sk[87]: Among the different initiatives of the Slovak National Centre for Quantum Technologies (QUTE.sk), the project EduQUTE[88] cite has the mission to create an interdisciplinary combined master program and expanding existing engineering programs through the inclusion of QT. Alongisde, it is also establishing a doctoral research center and providing support for postdoctoral research.